\relax
\documentstyle [prb,aps,preprint]{revtex}
\begin{document}
\title{\large\bf  Electron Locking 
In Semiconductor Superlattices\\}
\author{\large F.V.Kusmartsev \\}
\address{\normalsize \it Department of Physics, 
Loughborough University, LE11 3TU, UK \\
  \it and Department of Physics, University of Illinois at
Urbana-Champaign, \\ 1110 West Green Street, Urbana, IL 61801 \\
\it   Landau Institute, Moscow, Russia }
\author{\large Harjinder Singh Dhillon \\}
\address{\normalsize \it Department of Physics, 
Loughborough University, LE11 3TU, UK \\}

\date{\today}
\maketitle
\normalsize

\begin{abstract}

We describe a novel state of electrons and phonons
 arising in semiconductor superlattices (SSL) due to 
 strong electron-phonon interactions.  These states 
are characterized by a 
localization of phonons and a self-trapping or locking
 of electrons in one or several 
quantum wells due to  
additional, deformational potential arising around these locking
wells in SSL. The effect is enhanced
  in a 
longitudinal magnetic field.

Using the tight-binding 
and adiabatic approximations the
 whole energy spectrum of the self-trapped states is found 
and accurate, analytic expressions are included 
 for strong electron-phonon coupling.
   Finally, we discuss possible experiments which may detect these 
predicted self-trapped states.

\end{abstract}

\vspace{7mm}
PACS numbers:  63.20.Kr, 74.80.Dm, 73.20.Dx, 79.60.Jv

\vfill


Modern technology allows materials and structures
with narrow bandwiths to be created.  
One example of such a structure is the semiconductor superlattice which 
can be manufactured in such a way that it can consist of
different numbers of layers of differing thicknesses.  In SSL 
consisting of say, for example, a sequence of layers of
 $GaAs$ and $Ga_{x}Al_{1-x}As$, 
an  electron miniband of width $t$ may be created, where $t$ is 
the overlapping integral for an electron localized in neighboring 
quantum wells.  The number of levels in this miniband depend on the 
number of layers the SSL consists of.  For example, if the superlattice 
consists of two quantum wells, there are only two levels associated 
with the symmetrical and antisymmetric wave functions.  For each 
energy level, the probability to find the electron in either well 
is equally distributed.  The same is true when there are many levels 
in the miniband.  However, the situation changes dramatically if electron-phonon 
interactions are considered.  The effect of this interaction may give 
rise to coupled electrons and phonons where the 
dimensionless electron-phonon coupling constant of this system 
is inversely proportional to $t$.  Since the miniband width, in SSL, is
between 
1 - 100 meV the electron-phonon coupling is 
strongly intensified in comparison with the conventional 
bandwidth which is about 1 eV.  Such an intensification in the 
electron-phonon interaction in narrow band structures 
may give rise to some novel, self-trapped 
states of the electron. 

This phenomenon of the self-trapping
of electrons and excitons in solids, originally predicted theoretically 
\cite{Peka,Rash1,Hols,Rash2,KusST}, has been observed and well 
studied in many materials
(see, for example, review \cite{Rash2}).  It arises, primarily, in low 
dimensional systems and in systems with a strong electron-phonon 
interaction.

 The SSL is a new system where 
the effective electron-phonon coupling may become large, especially
when a longitudinal magnetic field is applied.
This can give 
rise to the locking of an electron in some wells
with the creation of novel self-trapped states.
 The number of such states depend on how many layers the 
SSL consists of and how strong the electron-phonon interactions are, 
ie the narrowness of the miniband.  The structure of these states is 
associated with a localization pattern.  This pattern indicates 
where the electron is localized, ie, 
in which wells the electron is locked in.

The system can be described in
the framework of a 1D tight-binding model. If 
the interaction of the electron with acoustic phonons is 
considered then the Schr\"odinger 
equation takes the form:

\begin{equation}  
  -{{\Delta \psi_n} \over {2m_{\perp}}} - t\psi_{n-1} + 2t\psi_n -
t\psi_{n+1} 
  + D Q_n \psi_n = E\psi_n  \label{sh-eq}  
\end{equation}

\noindent where $\Delta =\partial^2/\partial x^2+\partial^2/\partial y^2 $ is
a two dimensional Laplacian associated with the electron motion
inside the planes of the quantum wells, $m_{\perp}$ is the effective, 
transverse mass of the electron,
 $\psi_n (x,y)$ is the wave function for an 
electron located in the $n^{th}¥$ quantum well
and $D$ is a constant of the deformation potential which depends on 
the material the SSL is made up of. 
Note that deformations along all 3 coordinate axes have been included 
because $Q_n= \nabla. {\bf u}= \partial_x  u_x + 
\partial_y  u_y + \partial_z  u_z$ where 
${\bf u_n} = (u_{nx},u_{ny},u_{nz})$  is a vector of atomic
displacements in the $n^{th}¥$ quantum well.  Thus, 
$Q_n$ includes deformations in the $x$, $y$ and $z$ directions.

Using the adiabatic approximation, in which the lattice
moves slowly in comparison with the electrons
(because of the 
large mass of atomes $M_a$ and small, effective
mass $m_e=\hbar^2/t a^2$ of the electrons), 
the kinetic energy of the lattice with
the  parameter $m_e/M_a \rightarrow 0$ is zero.
The adiabatic self-trapped states are associated with the stationary
 points of the adiabatic potential.
 In other words employing the adiabatic parameter (the ratio of electron and
 atomic masses $m_e/M_a<<1$)
 following Pekar \cite{Peka} we consider different {\bf static}
 deformations of
 the lattice and the localized electron states created by these
 deformations. Extreme (or critical)  points of this adiabatic surface are
 self-trapped states.

 Traditionally,
 adiabatic self-trapping  has only been studied in continuum models 
\cite{KusST}
 applicable  to  localized states with a large radius (in comparison to the
 lattice spacing), except, of course, the exciton self-trapping in rare gas
 solids \cite{Rash2}. The same
 is true for studies of the interaction of an electron with polar phonons
 (polarons), where the adiabatic
 Pekar type polarons have a large radius. 
 Anti-adiabatic (or non-adiabatic) polarons  with a small radius have
 also been intensively investigated \cite{Alex}.
 However, the anti-adiabatic limit is completely the opposite
 to the adiabatic limit discussed in our paper. 
In the adiabatic approximation, the adiabatic potential $J$, 
consisting of the electron energy $E$
and the elastic energy of the lattice $E_{el}=K \sum_{n}{ Q_{n}^{2}/ 2}$,
is given by

\begin{eqnarray}
  J &=& \int d^2x[\sum_n \frac  { |\nabla \psi_n |^2} {2 m_{\perp} }+ 
  t \sum_{n}| \psi_{n}-\psi_{n+1} |^{2}  \nonumber  \\
    & &  + D \sum_{n}Q_{n}|\psi_{n}|^{2} + K \sum_{n}{
Q_{n}^{2}\over 2}] 
      \label{genadpot1}  
\end{eqnarray}      

The lowest energy self-trapped states correspond to a 
 wave function which is homogeneous
in the transverse direction $\psi_n(x,y)=\psi_n$ and is a true solution
of the above equation. 
Then the equations describing the self-trapped states
are obtained by minimization of $J$ with respect to $\psi_n$ and $Q_n$
provided that $ \sum_n \int d^2x |\psi_n|^{2}=1$
(see, for example, \cite {Rash2,KusST}). 
Doing so, ie  by a minimization
of $J$ with respect to $Q_n$, gives
 $Q_n= -{D \over {K }}|\psi_{n}|^2$, where $K$ is the modulus of elasticity 
 of the 
lattice.  Substituting for $Q_n(\psi_n)$ in eq. (\ref{sh-eq}), 
(\ref{genadpot1}), respectively, 
gives the nonlinear Schr\"odinger equation (NSE)  
  
\begin{equation}  
   -\psi_{n-1}  + 2 \psi_n - \psi_{n+1} - c|\psi_n|^2 \psi_n  =  
    E\psi_n  \label{dns1}
\end{equation}

\noindent and the corresponding expression for the adiabatic potential

\begin{equation}
  J = \sum_{n}| \psi_{n} -\psi_{n+1}|^{2} 
      -  {c \over 2} \sum_{n} \mid \psi_{n}\mid^{4}
 \label{adpot1}
\end{equation}      
 
\noindent where $c = {D^2}/{t K}$ is the dimensionless electron-phonon 
coupling constant and 
$E$ and $J$ are measured in units of the miniband width $t$. 
The normalization condition is 
$\sum_{n} \mid \psi_{n} \mid ^{2} =1/L^2$, where $L^2$ is the area of the
transverse
plane of the SSL (assuming it is a square of side $L$).  
It is convenient to make the scaling transformation
$\psi_n \rightarrow \psi_n/L$. This transformation
does not alter the form of the equations but it does give the conventional 
normalization
condition for the new wave function $\sum_n \mid \psi_{n}^{2}¥¥ \mid =1$. 
 However, the coupling constant does change so that $c_{new}=c_{old}/L^2$.
  In a longitudinal
 magnetic field the continuum transverse 
spectrum of the miniband
is split into Landau levels.
 For strong magnetic fields only the first
Landau level is relevant.   The electron  on this level is localized
in a transverse direction in an area  of the order of $l^2_B$,
where $l_B$ is a magnetic
length, such that, $l^2_B=C \hbar /eB$ ($L$ and $l_B$ are measured
in units of the SSL period $d$).
Then the electron motion is only one dimensional 
along the direction of the magnetic field through the SSL.
 The coupling constant then transforms to 
$c_{new}\rightarrow c_{old}/l^2_B$.  Thus, {\bf with the increase of
the longitudinal magnetic field, the phonon coupling constant 
can be strongly enhanced}.

To illustrate, consider some examples having exact solutions.  
The simplest is a double well structure.  
For a large coupling constant, $c \gg 1$, 
there arises the trapping of the electron in one 
of the two wells.  The eigenvalue of this state is $E=2-c$ and the 
corresponding eigenvectors are: $\psi_1= 
{\sqrt{1 \pm \alpha} \over \sqrt{2}}$ and $
\psi_2= {\sqrt{1 \mp \alpha} \over \sqrt{2}}$ where 
 $\alpha=\sqrt{1-{A \over {c^2}}}$, where $A=16$ or $A=4$ for 
periodic (PBC) or open boundary conditions, respectively. 
  These states are two-fold 
degenerate.

Another straight-forward example, studied analytically, 
consists of three superlattice layers or three quantum wells.  
The ground state is characterized by the localization of the electron in 
a single quantum well in the same manner as in the double well structure.
For PBC with an 
increase in the value of $c$ this state very rapidly saturates to the 
limit $\{\psi_{1}, \psi_{2}, \psi_{3}\} \Rightarrow \{0,1,0\}$ or 
$\{0,0,1\}$ or $\{1,0,0\}$.  
These three states are degenerate due to the translational symmetry 
associated with PBC and corresponds to the  ground state
energy which  in the limit $c\rightarrow\infty$ takes the simple form
$E=2-c$, ie the same as the 
double well structure.   There is also an 
excited self-trapped state, 
which, for $c \gg 1$, has the (degenerate) structure 
$\{\psi_{1}, \psi_{2}, \psi_{3} \} 
\Rightarrow \{{1/ \sqrt{2}},0,{1 / \sqrt{2}}\}$ or 
$\{{1 / \sqrt{2}},{1 / \sqrt{2}},0\}$ or $\{0,{1 / \sqrt{2}},
{1 / \sqrt{2}}\}$.  This state corresponds to the 
eigenvalue $E = 1 - {c \over 2}$ and the region of this self-trapped 
localization 
is two neighboring  wells.  There is also another self-trapped 
state in this system, which arises for very large values of $c$.  This state 
has the 
eigenvalue $E = 3 - {c \over 2}$ and the associated structure has the 
form $\{\psi_{1}, \psi_{2}, \psi_{3}\} \Rightarrow \{{1 \over 
\sqrt{2}},{-1 \over \sqrt{2}},0\}$ or analogous cyclic permutations of 
this pattern.  An
 exact solution, valid for all values of $c$, has also been found, 
 however, it does 
not appear to have a simple form.  Consequently, the exact solutions
are presented graphically in Fig.1.
The self-trapping states 
emerge for values of the coupling constant, $c > 3$.  From this figure
it is clear that the solutions obtained in the limit $c \rightarrow \infty$
describe the spectrum of the triple quantum well structure very accurately.  
The first corrections to the presented eigenvectors are of order 
$1/c$.  

Without PBC the translational invariance is broken and
 the degeneracy is lifted so that the number of energy levels 
increases.  The exact spectrum for the triple quantum well structure, 
without PBC, is presented 
in Fig. 2.
These new levels may also be described analytically 
for $c \gg 1$.  
A new eigenvalue, $E = 2 - {c \over 2}$ appears, which is associated 
with the localized self-trapped 
solution $\{{1 \over \sqrt{2}}, 0, {\pm 1 \over \sqrt{2}}\}$.  
Additional eigenvalues, 
which also satisfy the NSE, 
include $E = {2 - c \over 3}, {6 - c \over 3}, 
{10 - c \over 3}$.  These eigenvalues are associated with 
states where the electron is localized in each of the quantum wells with 
 equal probabilities.  Again, 
for nearly all values of coupling constant $c$, the amazing coincidence 
between the spectra obtained for $c \gg 1$ and the exact 
result shown in Fig.2 is seen.

From the examples discussed above, it is noticed that 
in the limit $c \rightarrow \infty$, the spectrum has some 
special features associated with the pattern of the localized states.  
The spectrum depends on how large the set of 
localizing quantum wells is and how many spots (groups of 
neighboring, localizing quantum wells) this set is separated into.  
The lowest eigenvalue always has the form $E = 2 - c$ and is associated 
with the localization of the electron in a single quantum well.  The 
next eigenvalue, $E = {(2 - c) \over 2}$, is associated with the 
localization of the electron in a single spot consisting of
two neighboring quantum wells.  
However, when the localization occurs in two separate quantum wells, 
so that there are two spots each consisting of one well,  
the energy eigenvalue is $E = {(4 - c)  \over 2}$.  
With the localization of the electron in three quantum wells the 
associated eigenvalue may be written as $E = {2 - c \over 3}$.

This observation can be used to derive exact solutions 
associated with an arbitrary localization 
pattern of an electron in SSL consisting of $N$ quantum wells.  
Initially, in the zero approximation
with parameter $c \gg 1$, assume that  the electron is 
localized in a single quantum well of a $N-$well SSL, 
say the $k^{th}$ well, so that
$|\psi_{k}|^2 = 1$ and $|\psi_{n}|^2 = 0$ for 
all possible $n \neq k$.  Upon making this substitution into the NSE 
and after some algebraic manipulation, we see that this substitution 
corresponds to a solution
with the eigenvalue $E = 2 - c$.  Similarly, if it is assumed 
that the electron is 
localized in two neighboring quantum wells, say, the $k^{th}$ and
$(k+1)^{th}$, 
and we substitute into the NSE $\psi_{k} = {1 \over \sqrt{2}}$, 
$\psi_{k+1} = {1 \over \sqrt{2}}$ and $\psi_{n} = 0$ for $n \neq k, 
n \neq k+1$ then (after some algebra) 
the eigenvalue $E = {2 - c \over 2}$ is obtained.  For the 
localization of the electron in 3 neighboring 
wells, we make the substitution $\psi_{k} = \psi_{k+1} = \psi_{k+2} 
= {1 \over \sqrt{3}}$ and $\psi_{n} = 0$ for all other wells.  This 
then yields $E = {2 - c \over 3}$.  Repeating the same procedure for 
the electron localized in $n$ neighboring wells, we derive the 
eigenvalue $E = {2 - c \over n}$.

The same procedure may be applied for an electron localized in two separate 
spots, consisting of, say, $n_{1}^Ë$ and $n_{2}^Ë$ quantum 
wells.  Surprisingly, we find that in this case, 
$E = {(4 - c) \over n}$, where $n = n_{1}^Ë + n_{2}^Ë$.  If this 
localization occurs in $m$ distinct spots, and each spot consists 
of $n_{1}^Ë,n_{2}^Ë,\ldots,n_{m}^Ë$ neighboring quantum wells, the 
associated eigenvalue is $E = {(2m - c) \over n}$, where 
$n = n_{1}^Ë + n_{2}^Ë + \ldots + n_{m}^Ë$. 

However, this is still not a complete set of  solutions.  Note that if 
the electron is localized in a 
spot consisting of $n_{1}^Ë = 2  n_{0}^Ë$ neighboring 
quantum wells, then the wave function may change sign for the different 
quantum wells of this spot.  For example, if for the first 
$n_{0}^Ë$ neighboring quantum wells the wave 
function has a positive sign and for the remaining $n_{0}^Ë$ wells the wave 
function is negative, then the associated eigenvalue is 
$E = {2m + 4 - c \over n}$.  
In general, if the wave function changes sign $l$ times in one or 
several localized spots the energy eigenvalue of the NSE takes
the form

\begin{equation}
    E = {2m + 4l - c \over n}    \label{eieq}
\end{equation}

\noindent
where $n$ is the total number of quantum wells in which the 
electron is localized. Note that, in each localizing quantum well
 the electron
is localized with equal probability, $\psi_{k}^{2} = 1/n$ for all 
numbers $k$ associated with localizing quantum wells. This probability does
not
depend on the number of spots, $m$, 
the localization area
is separated into.    The number $n$ may take any integer 
value $n = 1,2,3,\ldots,N$.  
The number $m$ may take any integer value satisfying 
$m \leq {n}$.  
Similarly, the number $l$ satisfies $l \leq n-1$.

Eq. (\ref{eieq}), 
obtained in the limit $c \rightarrow \infty $, has been compared 
with the exact 
solutions for systems consisting of up to 5 quantum wells, in the same manner 
as for the triple quantum well structure.  Also a comparison has been 
made between energy 
spectra obtained from equation (\ref{eieq}) and corresponding eigenvectors  
and energy eigenvalues from numerical 
calculations for SSL consisting of up to 11 quantum wells.  
In all these cases for nearly all values of $c$ (except small regions
of critical values where the self-trapped solutions originate) 
there is perfect agreement
with the derived formula (\ref{eieq}).
 However, in contrast with this perfect agreement between
 eigenvalues, a decrease in the value of $c$ leads to a noticable
 deviation in the  
wave functions (eigenvectors)  from those  
 obtained in the limit $c\rightarrow\infty$.
The first order corrections to the presented eigenvectors obtained with
the use of perturbation theory is of the order
of $\sqrt{n}/c$. When the coupling constant $c$ is not very large
 the wave functions of some states
may have interesting incommensurate and
chaotic structures. The detailed
analysis of such wave functions  will be published elsewhere \cite{KusDK}.
Thus, from the comparison with numerical results
and  with perturbation theory we conclude, 
that even though the spectrum of the NSE for the system 
with a finite, arbitrary number of wells $N$ 
is  well described by equation (\ref{eieq}) the shape of appropriate 
wave functions for smaller values of $c$ may have 
only qualitative features of the appropriate eigenvectors 
obtained in the limit $c \rightarrow \infty$.  
As $c$ decreases the localization spots smear out and the boundaries 
between localizing quantum wells and quantum wells where the wave 
function vanishes diminish.

In general, for large ranges of the values of $c$ the 
eigenvalue $E_{lmn}^Ë$ corresponds to the state of an electron 
which is self-trapped or localized with near equal probability 
in $n$ wells 
which are separated into $m$ groups (spots).  
Between the spots 
the electron wave function is nearly vanishing while 
inside these spots  the wave function changes sign $l$ times, 
although its amplitude is nearly the same.  
Since the spectrum, eq.(\ref{eieq}), is associated with a local
localization pattern, it is universal and 
does not depend on the boundary conditions.

The eigenvalues $E_{nml}^Ë$ are electron energies created in SSL by 
local deformations.  Therefore, it is interesting to estimate
an adiabatic potential,
$J_{nml}^Ë = E_{nml}^Ë + E_{el}^Ë$,
needed to create these states. 
With the use of the method described above  the following 
expression for the adiabatic potential is obtained,

\begin{equation}
     J_{nml}^Ë = {4m + 8l - c \over 2n}   \label{adpoteq}
\end{equation}

For comparison, a state associated with the bottom
of the miniband shifted by  electron-phonon interaction
corresponds to an adiabatic potential $J_{bs}=-c/(2N)$.
For a single localization spot with no change in the sign 
of the wave function, 
$l=0$ and $m=1$ are substituted into (\ref{adpoteq}).  This corresponds 
to the lowest value of adiabatic potential for all values of $n$. 
From these equations one sees that for strong
coupling, $c \gg 1$, there are
always  self-trapped states 
which have a lower adiabatic potential than
analogous miniband states, ie $J_{110}<J_{bs}$.
However, the states with very wide spots may have
$J_{n10}>J_{bs}$. This will occur when the size of the spot, $n$, 
becomes larger than some critical value $n_c =  N (c-4)/c$.  
The presence of these states (with 
$J_{n10}>J_{bs}$) indicates that the self-trapped states 
associated with 
the quantum numbers $n < n_c$ and the band states 
are separated by a self-trapped barrier, which appears only because 
of the lattice discreteness and finite size of the system.  
The self-trapping arises for a critical coupling equal to
$c_{crit}=4N/(N-1)$.  
In the continuum 1D  case the self-trapped barrier is absent although 
the critical coupling for the self-trapping does exist and 
depends on the size of the system \cite{Rash3}.  
The self-trapped barrier exists only in the continuum 3D case, as was 
first discovered by Rashba \cite{Rash1,Rash2}.

Thus, we have described a new phenomenon of
the self-organized, deformational creation of single,
double, triple and in general, multiple quantum  well structures in SSL
with an electron locked inside.
This is a local effect which depends on how many wells 
the SSL consists of (but exists for any number of wells) 
and on  how perfect the SSL is.  It may, therefore, be readily observed
in experiments. 
The states with the spectrum as described by equation (\ref{eieq}) 
may be detected, for example, by resonant tunneling 
experiments or by absorption of light.  Each self-trapped 
state associated with different quantum numbers $n, m, l$ 
will give rise to an 
absorption peak which lies below the miniband and
is well described by equation 
(\ref{eieq}).
However, as the self-trapped states are associated with 
deformations, which 
maintain the localization, the main absorption of light (which 
follows the Franck-Condon principle) will be associated with 
the miniband.  Since the creation of the self-trapped 
states takes less energy 
than the creation of the band states (see equation (\ref{adpoteq})), 
after absorbing radiation, the electrons excited at the absorption
of light inside the miniband then may be self-trapped.
  The emission of radiation from these states will 
directly indicate their presence.  The creation of excitons from 
these states is an interesting issue.

Another interesting feature of the self-trapped (or locked) states 
in SSL is 
their strong interaction with phonons or with ordinary 
sound waves. 
Each of the localizing spots of a self-trapped state 
is created by deformation and therefore
also associated with the phonon localisation. 
Since inside the spot the deformation $\sim 1/n$ and vanishes outside 
then each spot
may be viewed as a sound wave resonator.
The resonant frequency and the wavelength of this resonator
is related to the size of the spot. For example, for the spot consisting
of $n_i$ quantum wells the size
 is $n_i d$ (note that $d$ is a period of SSL and $\sum_i n_i =n$). Then 
 the resonance wavelength will be $\lambda= n_i d$ and the
 phonons with the wavevector $2 \pi /(n_i d)$
  will be especially strongly
 scattered or trapped by the localizing spot.
 Based on this fact one may suggest the following scenario for
  experiments.
 When a sound wave moving through a SSL  passes the 
self-trapped spot 
there will be a transfer of momentum from the sound wave to the spot.  
As a result of this scattering, the acoustic 
phonons with wave vector $q=2\pi/ (n_i d)$ 
will be absorbed 
forcing the self-trapped spot  to propagate 
through the lattice.
 Such motion may give 
rise to the creation of current induced by sound.  

Alternatively, if a bias voltage is applied to the SSL, 
creating a current, 
 a constant backflow of
appropriate phonons will be created. This backflow of selective
phonons will be especially strong at some bias voltages equal
to the interlevel spacing. Therefore, they may be observed  as  some minima
in the current voltage characteristics of the SSL in a high magnetic field.
This effect may exist only at low temperatures when there are no
current carriers in the miniband.  Otherwise, 
in doped SSL or at high temperatures 
the transport will be of the
Bloch type. Since the number of self-trapped states increases
with the total number of quantum wells in the SSL it is 
expected that the proposed
effect will be stronger in SSL with larger numbers of quantum wells.
Since the described  effect of the locking of an electron in quantum wells
 is caused by acoustical
phonons it is universal  and should not depend on what material
the SSL consists of. Recently we have become aware that a similar effect
has been seen by L. Eaves \cite{Eave} group.

In summary we have described novel, self-trapped states
of an electron associated with the self-organized, deformational creation
of  single, double  and  multiple quantum  wells which lock
the electron in SSL.
The exact energy spectrum associated with these states 
for a strong electron-phonon interaction, $D^2/(Kt) \gg 1$,
or for a strong longitudinal magnetic field is derived.  
The novel states associated with local deformations
 have very different properties from 
the conventional band states and therefore may give rise to some 
new effects like, for example, the creation of an electric current 
by sound.  
Alternatively, such states stimulate 
sound absorption and prevent phonon propagation.  This 
may give rise to unusual thermopower in SSL.

\noindent {\large \bf Figure Caption}\\

\noindent {\bf Fig. 1} The dependence of energy spectrum 
on the electron-phonon coupling constant $c$ for 
a triple quantum well structure obtained assuming PBC.  The lines of the 
spectrum correspond to (from bottom up on left hand side) the eigenvalues 
$E = 2 - c$, $E = (2 - c)/2$, $E = (6 - c)/2$, $E = -c/3$ and $E = (8 - c)/3$.

\noindent {\bf Fig. 2} The same as in Fig.1  for a triple quantum well 
structure obtained without assumption of PBC.  The lines of the spectrum 
correspond to (from bottom up on LHS) the eigenvalues $E = 2 - c$, 
$E = (2 - c)/2$, $E = (4 - c)/2$, $E = (6 - c)/2$, $E = (2 - c)/3$, 
$E = (6 - c)/3$ and $E = (10 - c)/3$.


\begin{thebibliography} {99}



 
\bibitem {Peka} S.I. Pekar,  
Untersuchungen \"uber die Electronentheorie  
des Kristallen,  Akedemie Verlag, Berlin, 1954.  

\bibitem {Rash1} E.I. Rashba, Opt. Spectr. {\bf 2}, 78, {\bf 2}, 88 (1957)  
  
\bibitem {Hols} T. Holstein, Ann. Phys. {\bf 8}, 343 (1959)  
 
 

\bibitem {Rash2}  E.I. Rashba, in: Excitons, ed. by E.I.Rashba and
M.D. Sturge, North-Holland (Amsterdam) 1982, p.543.

\bibitem{KusST} F.V.Kusmartsev, Phys.Rev. B43, (1991), 1345.
 F.V.Kusmartsev et al, Europhys. Lett. {\bf 42} 547 (1998)

\bibitem{Alex} A.S.Alexandrov and N.F.Mott, Polarons and Bipolarons (WS, 
Singapore) 1995

\bibitem{KusDK} H.S. Dhillon, F. V. Kusmartsev,  and K. E. K\"urten, 
in preparation and cf, also, in
 F. V. Kusmartsev and K. E. K\"urten, {\it Effects of Chaos in Quantum
Lattice Systems} in {\it Lecture Notes in Physics}, Vol.\ 284 (1997), edited
by J. W. Clark and M. L. Ristig, Springer-Verlag, NY-Heidelberg Berlin 


\bibitem{Rash3} E.I. Rashba,  Synth. Met. {\bf 64}, 255 (1994)

\bibitem{Eave} R. Hayden, L. Eaves et al,(private communications)
September 1998

\end{thebibliography}
\end{document}